\documentclass{ws-mplb}
\usepackage{graphicx,float}
\usepackage{dcolumn}
\usepackage{bm}
\usepackage{amsmath}
\usepackage[latin1]{inputenc}
\usepackage{psfrag}
\usepackage{epsfig}
\usepackage{slashed}
\usepackage{color}

\begin{document}

\markboth{Thiago Prud\^encio}{Cutoff independent RG flow equations for two-coupled chains model}

%
\catchline{}{}{}{}{}
%

\title{Cutoff independent RG flow equations for two-coupled chains model}

\author{Thiago Prud\^{e}ncio}

\address{Coordination in Science and Technology - CCCT,
Federal University of Maranh\~ao - UFMA, Campus Bacanga, 65080-805, S\~ao
Luis-MA, Brazil.\\
thprudencio@gmail.com}

\maketitle


\begin{abstract} 
One-dimensional strongly correlated electron systems coupled via transverse hopping and presence of interband interactions can converge to a Luttinger liquid state or diverge to an even more intricate behavior, as a Mott state. Explicit consideration of the renormalization group (RG) flow of the Fermi points in the Fermi surface, turns the classification of phase transitions more challenging. We reconsider the recent paper for the spinless case [E. Correa and A. Ferraz, Eur. Phys. J. B 87 (2014) 51], where RG flow equations are derived in a cutoff-dependent form up to two-loops order. We demonstrate that the cutoff-dependence can be removed by rewriting the RG flow equations in terms of the energy scale variable. In our paper, the RG flow equations assume a cutoff-independent form and leads to fixed points independent of cutoff choice. The consequence is the invariance under cutoff transformations, more suitable for classifying universality classes and phase transitions.
\end{abstract}


\section{Introduction}

\begin{figure}[t]
\centering
\includegraphics[scale=0.4]{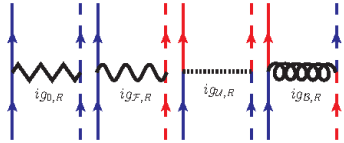}
\caption{(Color online) Band interactions on TCCM: forward $-ig_{0,R}$, interforward $-ig_{\mathcal{F},R}$, 
backscattering $-ig_{\mathcal{B},R}$ and umklapp $-ig_{\mathcal{U},R}$.} 
\label{interacoesTCM}
\end{figure}
\begin{figure}[b]
\centering
\includegraphics[scale=0.23]{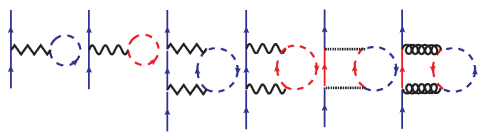}
\caption{(Color online) $\Gamma_{+,R}^{a(2)}$ diagrams up to 2-loops.}
\label{selfenergyb}
\end{figure}
\begin{figure}[h]
\centering
\includegraphics[scale=0.23]{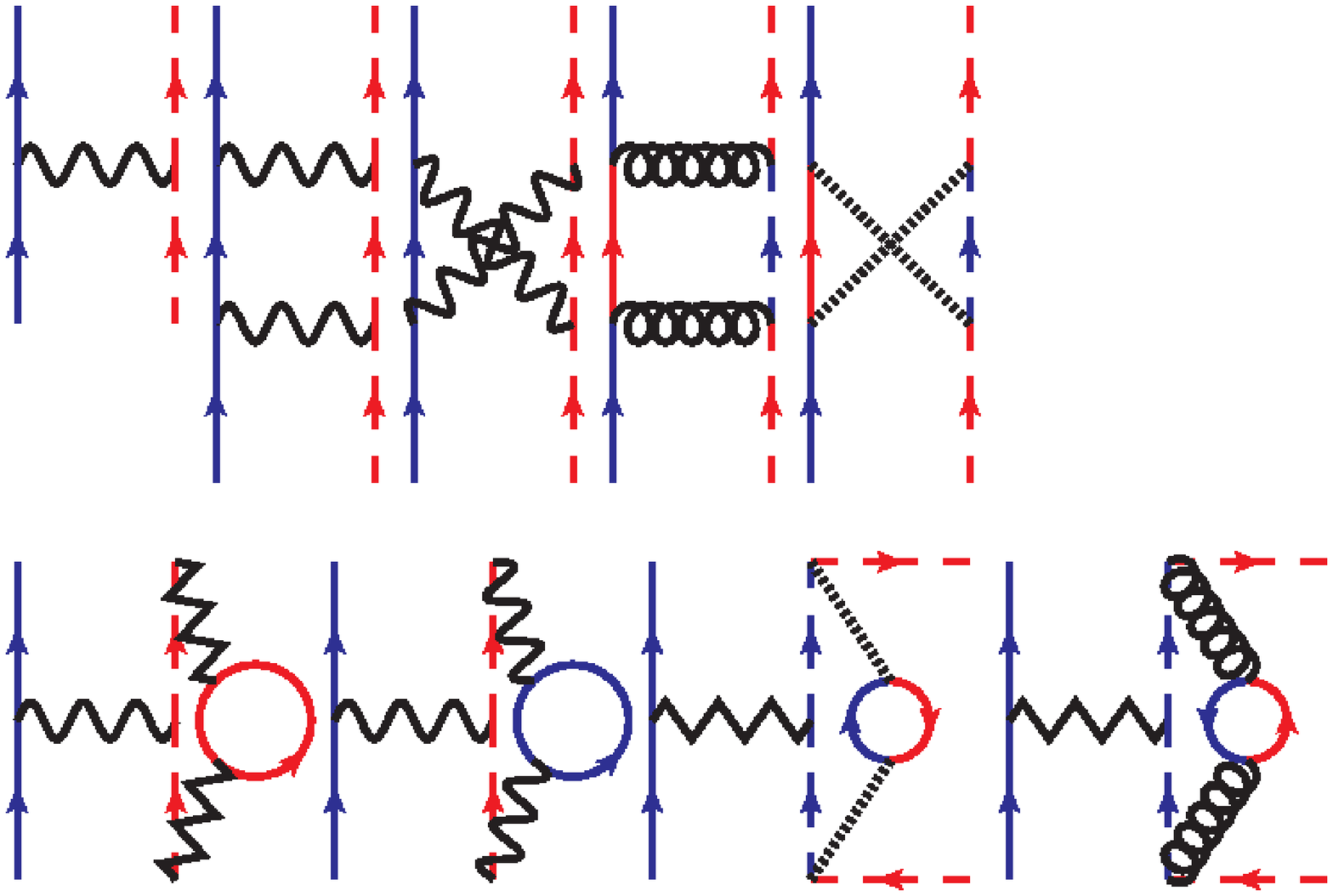}
\includegraphics[scale=0.23]{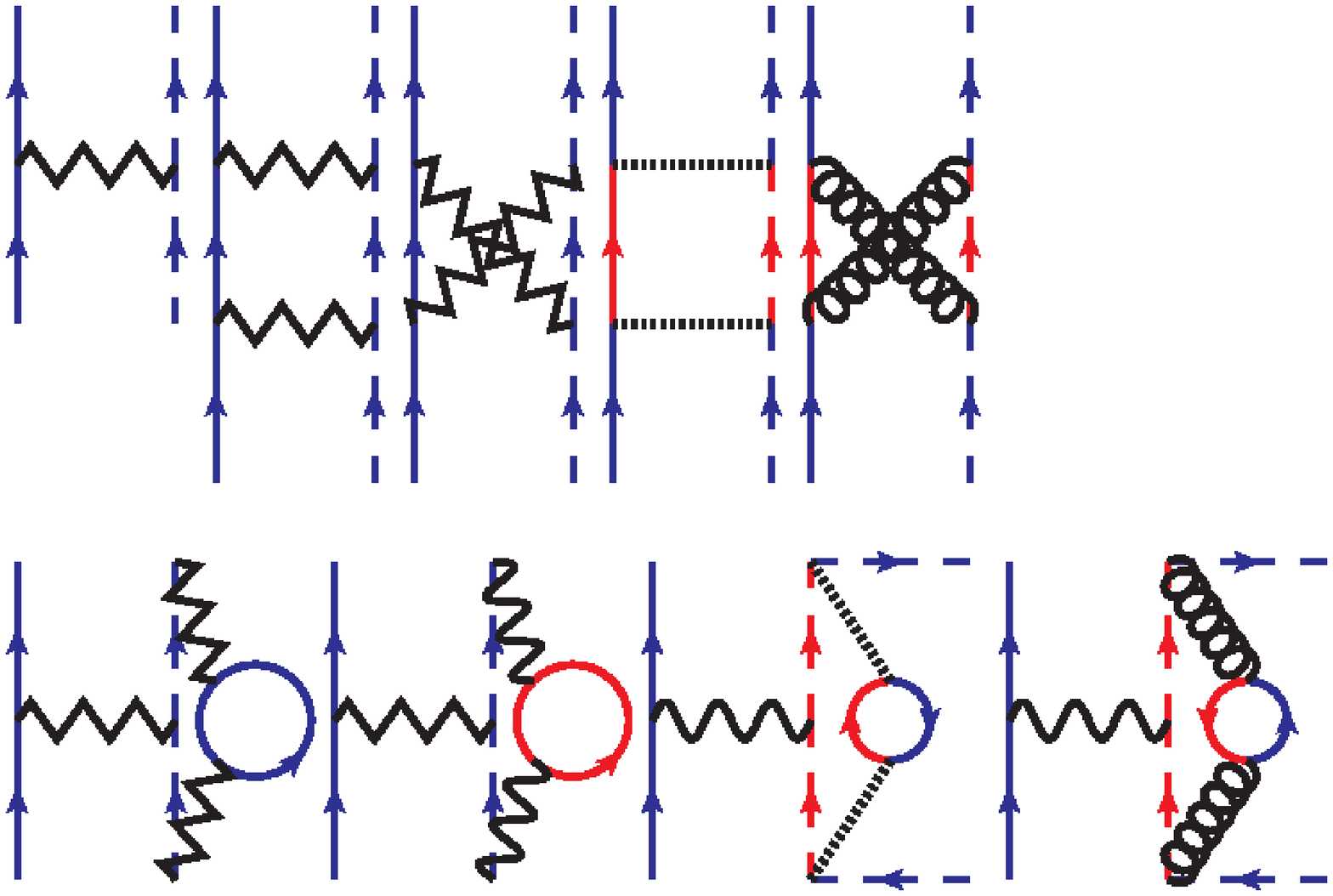}
\includegraphics[scale=0.23]{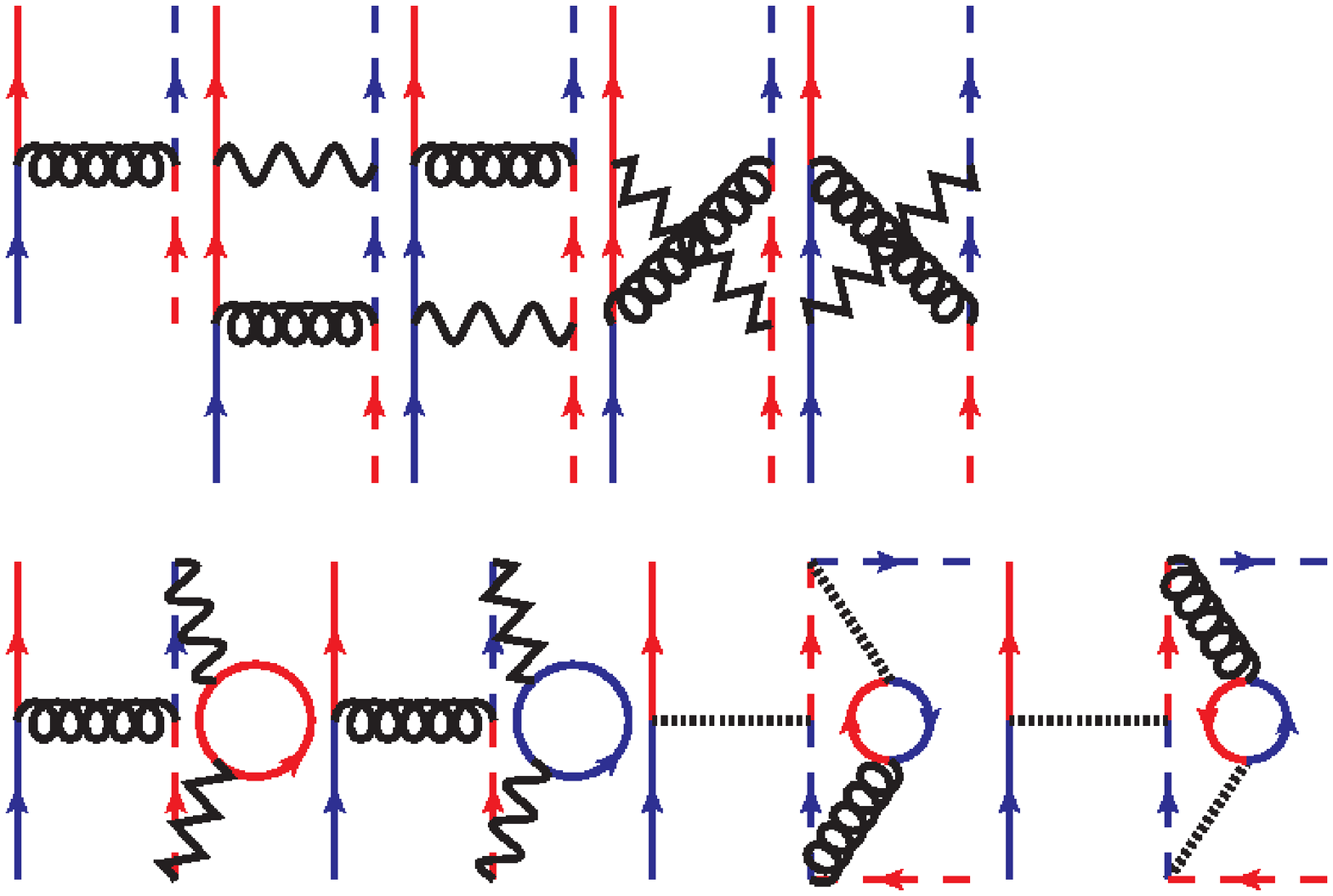}
\includegraphics[scale=0.23]{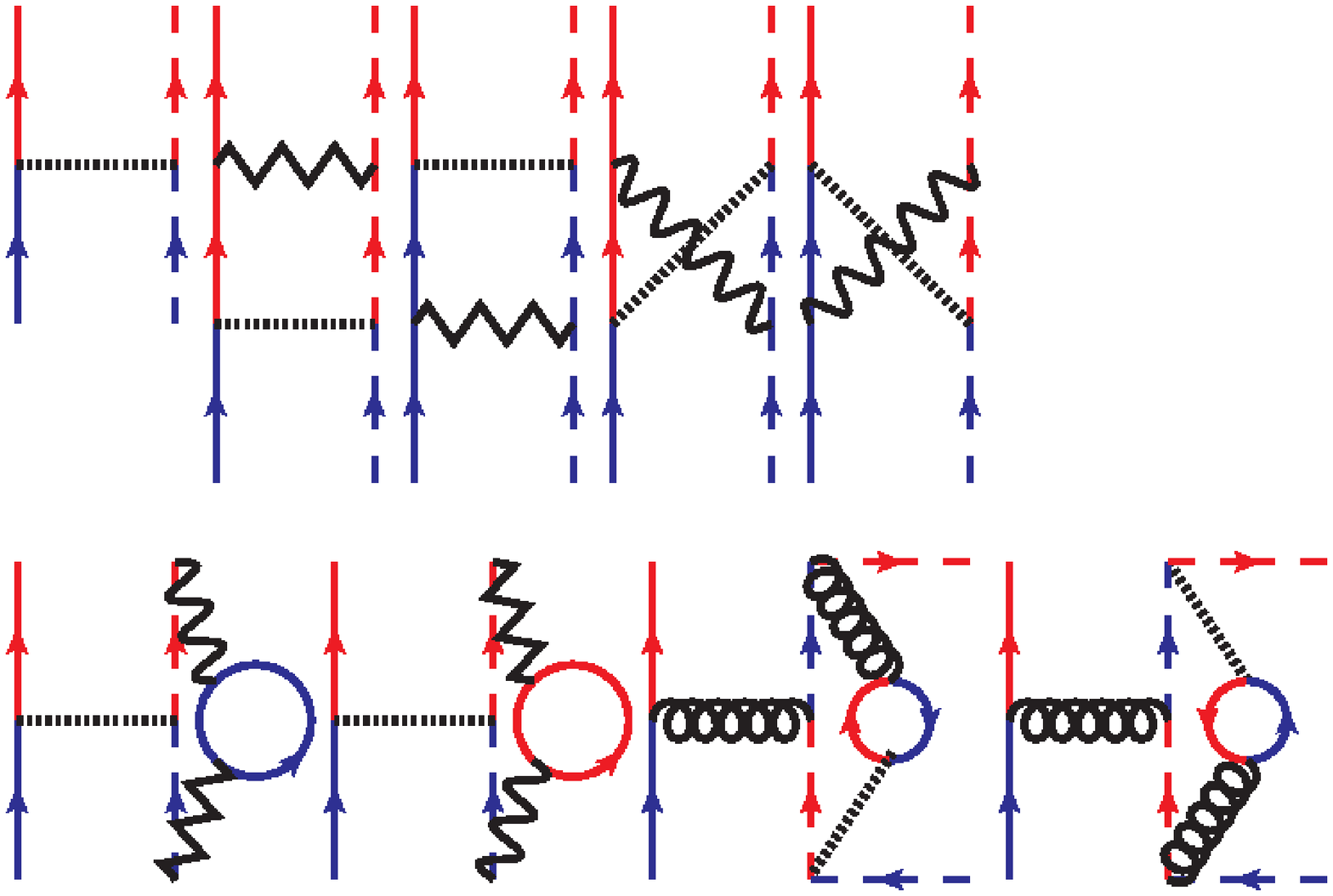}
\caption{(Color online) $\Gamma^{\alpha(4)}_{i,R}(\lbrace{\bf{k}}_{i}\rbrace)$ diagrams up to 2-loops, for $i=0,\,\mathcal{F},\,\mathcal{B},\,\mathcal{U}$.}
\label{gamma4}
\end{figure}

When one dimensional strongly correlated electron systems are coupled via transverse hopping in the presence of 
interband interactions, a Luttinger liquid state or another more exotic phase can emerge, as a Mott insulating state 
\cite{yoshida}. Taking into account explicitly the renormalization 
group (RG) flow of Fermi points forming the Fermi surface (FS) turns the classification of phase transitions 
more challenging. In the spinless case, the influence 
represented by the transverse 
hopping, $t_{\perp}$, in the absence of interactions lead to the model of two coupled 1D spinless chains 
\cite{fabrizio,ledermann}, presenting a richer behaviour when intraband and interband interactions are included 
\cite{fabrizio}. In this form the so called two coupled chains model (TCCM) represents a prototype for the 
study confinement as proposed by Anderson \cite{anderson}.  

In the recent RG analysis up two-loops order of the two coupled chains model (TCCM) \cite{eberth} has exhibited a 
set cutoff dependent RG flow equations. This problem has also been discussed in association with quantum anomaly 
\cite{prudencio1,alvaro2}. The contrast of these works with a previous one by Fabrizio \cite{fabrizio} was 
the reconsideration of the renormalization of Fermi points associated to the two bands appearing in the TCCM. This new point 
of view turns this problem richer and capable of describing physically the effects of 
FS reconstruction under the electronic interactions that renormalizes the FS. When the set of RG flow equations is 
derived by field renormalization group (RG) in the TCCM, the presence of scattering channels of intraband forward, 
interband forward, interband backscattering and interband umklapp (figure \ref{interacoesTCM}) up two-loop orders produces contributions 
for the renormalized selfenergy (figure \ref{selfenergyb}) and renormalized scattering channels up to two-loops order 
(figure \ref{gamma4}). The renormalization of Fermi points also brings new features associated to the presence of possible anomaly 
contributions \cite{alvaro2}. 

We reconsider the recent propose RG for TCCM \cite{eberth} where RG flow equations are 
derived in a cutoff dependent form up two-loops order. The presence of cutoff dependence in the renormalization group 
(RG) flow equations is, however, problematic, 
since for a consistent RG method the RG invariance has to be part of the RG flow equations in order to have an universality 
behaviour \cite{collins}. When a set of RG flow equations is independent of the energy scale cutoff, the 
corresponding fixed points are not restricted to a given cutoff choice. Although the cutoff dependence in the case proposed, 
as we will show, is associated 
to a given choice of parameter, and consequently can be removed by reconsidering a more apropriate parameter. 
For the parameter proposed for the RG flow, the solutions can be changed by changing the choice of the cutoff 
contribution. 

Here we demonstrate that this cutoff dependence can be removed by rewriting the RG flow equations in terms of the 
energy scale variable $\omega$. In our propose, the RG flow equations assume a cutoff independent form and consequently the 
fixed points will be independent of cutoff choice. The consequence of this new form is the invariance of the RG flow under 
cutoff transformations, turning the problem more suitable for classification of universality classes. The RG flow equations then 
turn to be consistent with the RG methods with cutoff independent RG flow equations.


\section{RG flow equations with cutoff dependence}
By considering the RG of the TCCM with the consideration of two-loop order contribution in the scattering channels 
of intraband forward, 
interband forward, interband backscattering and interband umklapp (figure \ref{interacoesTCM}), the renormalized 
selfenergy (figure \ref{selfenergyb}) and renormalized scattering channels 
(figure \ref{gamma4}), will give rise to the following set of RG flow equations for the TCCM, that when written written 
in terms of $l$ variable, $l=\ln\left(\frac{\Omega}{\omega}\right)$,
take the form \cite{eberth}  
\begin{eqnarray}
\frac{d\bar{g}_{0R}}{dl} &=& -\bar{g}_{0R}M_{0R} + \bar{g}_{\mathcal{F}R}N_{0R} \nonumber \\
&-&\bar{g}_{\mathcal{U}R}^{2} + \bar{g}_{\mathcal{B}R}^{2}C,\\
\frac{d\bar{g}_{\mathcal{F}R}}{dl} &=& -\bar{g}_{\mathcal{F}R} M_{\mathcal{F}R} + \bar{g}_{0R}N_{\mathcal{F}R} \nonumber \\
&+& \bar{g}_{\mathcal{U}R}^{2} -\bar{g}_{\mathcal{B}R}^{2}C,\\
\frac{d\bar{g}_{\mathcal{B}R}}{dl} &=& -\bar{g}_{\mathcal{B}R} M_{\mathcal{B}R},\\
\frac{d\bar{g}_{\mathcal{U}R}}{dl} &=& -\bar{g}_{\mathcal{U}R} M_{\mathcal{U}R},\\
\frac{d\Delta k_{F,R}}{dl} &=& -\Delta k_{F,R}\bar{g}_{\mathcal{B}R}^{2}C,\\
\frac{dZ}{dl} &=& -Z \gamma,
\end{eqnarray}
where 
\begin{eqnarray}
M_{0R}&=& 2\gamma -\left(\bar{g}_{\mathcal{F},R}^{2} + \bar{g}_{0,R}^{2}\right)  \\
N_{0R}&=& \bar{g}_{\mathcal{U},R}^{2} + \bar{g}_{\mathcal{B},R}^{2}C  \\
M_{\mathcal{F}R}&=&  2\gamma -\left(\bar{g}_{\mathcal{F},R}^{2} + \bar{g}_{0,R}^{2}\right)  \\
N_{\mathcal{F}R}&=& \bar{g}_{\mathcal{U},R}^{2} + \bar{g}_{\mathcal{B},R}^{2}C \\
M_{\mathcal{B}R}&=& 2\gamma + \left(\bar{g}_{\mathcal{F},R} + \bar{g}_{0,R} - \bar{g}_{\mathcal{U},R}^{2}\right)(1+C) \nonumber \\
&+& 2\bar{g}_{\mathcal{F},R}\bar{g}_{0,R}C,\\
M_{\mathcal{U}R}&=& 2\gamma + 2\left(\bar{g}_{\mathcal{F},R} + \bar{g}_{0,R}\right) - \bar{g}_{\mathcal{B},R}^{2}C \nonumber \\
&-& 2\bar{g}_{\mathcal{F},R}\bar{g}_{0,R},\\
 \gamma &=& \frac{1}{2}\left(\bar{g}_{0,R}^{2} + \bar{g}_{\mathcal{F},R}^{2} + \bar{g}_{\mathcal{U},R}^{2} + \bar{g}_{\mathcal{B},R}^{2}C  \right),
\end{eqnarray}
and 
\begin{eqnarray}
C&=&C(\Lambda_{0}, l, \Delta k_{F,R})\nonumber\\
&=&\frac{1}{2}[\frac{1}{1 + \frac{\Delta k_{F,R}}{\Lambda_{0}}e^{l}} +
\frac{sign(1 - \frac{\Delta k_{F,R}}{\Lambda_{0}}e^{l})}{|1 - \frac{\Delta k_{F,R}}{\Lambda_{0}}e^{l}|}].\label{eqrg}
\end{eqnarray}
is a function with the presence of cutoff contribution to be discussed in the next section. 

It can also be observed that $M_{0R}=M_{\mathcal{F}R}$ and $N_{0R}=N_{\mathcal{F}R}$.

\section{Contribution of the cutoff term}

When the band Fermi velocity are equal $v^{a}_{F}=v^{b}_{F}=v_{F}$, the associated cutoff term $\Lambda_{0}$ leads to an 
energy cutoff written in the form 
\begin{eqnarray}
\Omega=2v_{F}\Lambda_{0}.
\end{eqnarray}
The energy cutoff and the energy scale are related by the following relation,
\begin{eqnarray}
\omega=\Omega e^{-l}=2v_{F}\Lambda_{0}e^{-l}, \label{op}
\end{eqnarray}
resulting that $e^{-l}=\omega/2v_{F}\Lambda_{0}$ or, equivalently,
\begin{eqnarray}
e^{l}=\frac{2v_{F}\Lambda_{0}}{\omega}. \label{1}
\end{eqnarray}
With this expression we can write the explicit form of the $l$ variable, responsible for the logarithmic divergences and
 given by 
\begin{eqnarray}
l=\ln\left(\frac{\Omega}{\omega}\right).
\end{eqnarray}
One consequence of this relation is that the infinitesimal term is independent of cutoff
\begin{eqnarray}
dl &=& -\frac{d\omega}{\omega}. \label{if8}
\end{eqnarray}
The RG flow equations derived in \cite{eberth} carry the cutoff term $\Lambda_{0}$ due to the presence of the following 
cutoff dependent contribution
\begin{eqnarray}
\frac{\Delta k_{F,R}}{\Lambda_{0}}e^{l}.
\end{eqnarray}
This cutoff term is present if we write in terms of $l$ variable for the RG flow, but can be eliminated if we write in terms of 
the energy scale variable $\omega$ as follows 
\begin{eqnarray}
\frac{\Delta k_{F,R}}{\Lambda_{0}}e^{l}=\frac{\Delta k_{F,R}}{\Lambda_{0}}\frac{2v_{F}\Lambda_{0}}{\omega}=\frac{2v_{F}\Delta k_{F,R}}{\omega}.
\end{eqnarray}
As such, by writing the energy scale in terms of $l$, we have an explicit cutoff dependence in such a way that a contribution 
due to both cutoff choice and $l$ variable will be present
\begin{eqnarray}
\omega =\omega(\Lambda_{0},l) = 2v_{F}\Lambda_{0}e^{-l},
\end{eqnarray}
and the rate of change with the cutoff choice can be given by the derivatives
\begin{eqnarray}
\frac{\partial \omega(\Lambda_{0},l)}{\partial \Lambda_{0}}&=&2v_{F}e^{-l},\\
\frac{\partial \omega(\Lambda_{0},l)}{\partial l}&=& -2v_{F}\Lambda_{0}e^{-l}.
\end{eqnarray}
The cutoff dependent contribution that appears in the the RG flow equations for the TCCM is given by the equation (\ref{eqrg})
\begin{eqnarray}
C(\Lambda_{0}, l, \Delta k_{F,R})&=&\frac{1}{2}[\frac{1}{1 + \frac{\Delta k_{F,R}}{\Lambda_{0}}e^{l}} +
\frac{sign(1 - \frac{\Delta k_{F,R}}{\Lambda_{0}}e^{l})}{|1 - \frac{\Delta k_{F,R}}{\Lambda_{0}}e^{l}|}]. \nonumber
\end{eqnarray}
This form has an explicit dependence of the cutoff term $\Lambda_{0}$. 

Since we can also write 
\begin{eqnarray}
1\pm \frac{\Delta k_{F,R}}{\Lambda_{0}}e^{l}= 1 \pm \frac{2v_{F}\Delta k_{F,R}}{\omega},
\end{eqnarray}
this term is indeed cutoff independent when written in terms of the $\omega$ variable. 

We can then write 
$C(\Lambda_{0}, l, \Delta k_{F,R})=C(\omega, 2v_{F}\Delta k_{F,R})$, where
\begin{eqnarray}
C(\omega, 2v_{F}\Delta k_{F,R})&=& \frac{1}{2}[\frac{1}{1 + \frac{2v_{F}\Delta k_{F,R}}{\omega}} +
\frac{sign(1 - \frac{2v_{F}\Delta k_{F,R}}{\omega})}{|1 - \frac{2v_{F}\Delta k_{F,R}}{\omega}|}],
\end{eqnarray}
that is a completelly independent of cutoff term, in terms of the energy scale variable $\omega$.

\section{The numerical fixed points}

For a given $l$ and two different cutoffs, for instance $\Lambda_{0}=10^{3}$ or $\Lambda_{0}=10^{2}$, two different 
energy scales will be associated to the RG flow. As a consequence the cutoff term will have influence in 
the RG flow in terms of $l$.  

The profile of RG flow is changed in terms of the $l$ variable, if, for instance, the choice $\Lambda_{0}=10^{3}$ is changed to 
the cutoff $\Lambda_{0}=10^{2}$. This implies that the choice of the $l$ variable instead of the energy scale $\omega$ is not 
a completelly consistent RG procedure since it will lead to RG flow equations whose solutions are restricted to a given cutoff, 
i. e., the flow equations will depend 
of the cutoff choice and will not be invariant under a cutoff transformation as $\Lambda_{0}=10^{3}\rightarrow \Lambda_{0}=10^{2}$. 

Since the RG flow equations in 
terms of $\omega$ do not carry a cutoff term, we have a consistent RG procedure in terms of this 
variable choice, whose profile of the RG flows stay invariant under the transformation 
$\Lambda_{0}=10^{3}\rightarrow \Lambda_{0}=10^{2}$.  

In order to make this point more clear, let us consider the way the numerical fixed points are affected by 
the choice of $l$ variable or $\omega$ variable. 

In terms of the $l$ variable, the choice of cutoff will interfere in the way in which the RG flow equations for the TCCM 
achieve a fixed point (It is not cutoff invariant). For instance, the choices of cutoff $10$, $10^{2}$ and $10^{3}$ will 
lead to three different equations determined by the different functions of cutoff contribution, given by
\begin{eqnarray}
C(10, l, \Delta k_{F,R})&=&\frac{1}{2}[\frac{1}{1 + \frac{\Delta k_{F,R}}{10}e^{l}} +
\frac{sign(1 - \frac{\Delta k_{F,R}}{10}e^{l})}{|1 - \frac{\Delta k_{F,R}}{10}e^{l}|}], 
\end{eqnarray}
\begin{eqnarray}
C(10^{2}, l, \Delta k_{F,R})&=&\frac{1}{2}[\frac{1}{1 + \frac{\Delta k_{F,R}}{100}e^{l}} +
\frac{sign(1 - \frac{\Delta k_{F,R}}{100}e^{l})}{|1 - \frac{\Delta k_{F,R}}{100}e^{l}|}],
\end{eqnarray}
and
\begin{eqnarray}
C(10^{3}, l, \Delta k_{F,R})&=&\frac{1}{2}[\frac{1}{1 + \frac{\Delta k_{F,R}}{1000}e^{l}} +
\frac{sign(1 - \frac{\Delta k_{F,R}}{1000}e^{l})}{|1 - \frac{\Delta k_{F,R}}{1000}e^{l}|}],
\end{eqnarray}
respectively.

Three choices of cutoff lead to three different RG equations in terms of $l$. As a consequence, 
the numerical fixed points in the solution of the RG flow equations in the $l$ variable will lead to different 
fixed points to each choice of cutoff.

Now let us consider the choices of cutoff $10$, $10^{2}$ and $10^{3}$ when writting the RG flow equation in the 
energy scale variable $\omega$. Since in this configuration, the function of the cutoff contribution takes the form
\begin{eqnarray}
C(\omega, 2v_{F}\Delta k_{F,R})&=& \frac{1}{2}[\frac{1}{1 + \frac{2v_{F}\Delta k_{F,R}}{\omega}} +
\frac{sign(1 - \frac{2v_{F}\Delta k_{F,R}}{\omega})}{|1 - \frac{2v_{F}\Delta k_{F,R}}{\omega}|}],
\end{eqnarray}
the three cutoff choices $10$, $10^{2}$ and $10^{3}$ will lead to the same RG flow equations when written in terms of energy scale 
variable $\omega$. 

The RG equations in terms of energy scale variable $\omega$ are then cutoff invariance, instead of in term os the $l$ variable. The 
fixed points are not equivalent.

In general, given a cutoff transformation $\Lambda_{0} \rightarrow \tilde{\Lambda}_{0}$, the RG equations in terms of the $l$ variable will change 
its form, while the RG equations in terms of $\omega$ will mantain the same form, carrying an universality property for the RG flow 
equations and consequently with an unique fixed point profile independent of the cutoff choice.

\section{RG flow equations independent of cutoff}

In terms of the energy scale variable $\omega$, the set of RG flow equations 
for the TCCM can rewriten as follows
\begin{eqnarray}
-\omega \frac{d\bar{g}_{0R}}{d\omega} &=& -\bar{g}_{0R}M_{0R} + \bar{g}_{\mathcal{F}R}N_{0R} \nonumber \\
&-&\bar{g}_{\mathcal{U}R}^{2} + \bar{g}_{\mathcal{B}R}^{2}C,\\
-\omega \frac{d\bar{g}_{\mathcal{F}R}}{d\omega} &=& -\bar{g}_{\mathcal{F}R} M_{\mathcal{F}R} + \bar{g}_{0R}N_{\mathcal{F}R} \nonumber \\
&+& \bar{g}_{\mathcal{U}R}^{2} -\bar{g}_{\mathcal{B}R}^{2}C,\\
-\omega \frac{d\bar{g}_{\mathcal{B}R}}{d\omega} &=& -\bar{g}_{\mathcal{B}R} M_{\mathcal{B}R},\\
-\omega \frac{d\bar{g}_{\mathcal{U}R}}{d\omega} &=& -\bar{g}_{\mathcal{U}R} M_{\mathcal{U}R},\\
-\omega \frac{d\Delta k_{F,R}}{d\omega} &=& -\Delta k_{F,R}\bar{g}_{\mathcal{B}R}^{2}C,\\
-\omega \frac{dZ}{d\omega} &=& -Z \gamma,
\end{eqnarray}
where 
\begin{eqnarray}
M_{0R}&=& M_{\mathcal{F}R}= 2\gamma -\left(\bar{g}_{\mathcal{F},R}^{2} + \bar{g}_{0,R}^{2}\right)  \\
N_{0R}&=& N_{\mathcal{F}R}= \bar{g}_{\mathcal{U},R}^{2} + \bar{g}_{\mathcal{B},R}^{2}C  \\
M_{\mathcal{B}R}&=& 2\gamma + \left(\bar{g}_{\mathcal{F},R} + \bar{g}_{0,R} - \bar{g}_{\mathcal{U},R}^{2}\right)(1+C) \nonumber \\
&+& 2\bar{g}_{\mathcal{F},R}\bar{g}_{0,R}C,\\
M_{\mathcal{U}R}&=& 2\gamma + 2\left(\bar{g}_{\mathcal{F},R} + \bar{g}_{0,R}\right) - \bar{g}_{\mathcal{B},R}^{2}C \nonumber \\
&-& 2\bar{g}_{\mathcal{F},R}\bar{g}_{0,R},\\
 \gamma &=& \frac{1}{2}\left(\bar{g}_{0,R}^{2} + \bar{g}_{\mathcal{F},R}^{2} + \bar{g}_{\mathcal{U},R}^{2} + \bar{g}_{\mathcal{B},R}^{2}C  \right),
\end{eqnarray}
and now the cutoff contribution $C$ is in fact cutoff independent 
\begin{eqnarray}
C(\omega, 2v_{F}\Delta k_{F,R})&=& \frac{1}{2}[\frac{1}{1 + \frac{2v_{F}\Delta k_{F,R}}{\omega}} +
\frac{sign(1 - \frac{2v_{F}\Delta k_{F,R}}{\omega})}{|1 - \frac{2v_{F}\Delta k_{F,R}}{\omega}|}]. \label{fuc}
\end{eqnarray}

\section{Cutoff contribution in the energy scale variable}

\begin{figure}[!htb]
\centering
\includegraphics[scale=0.5]{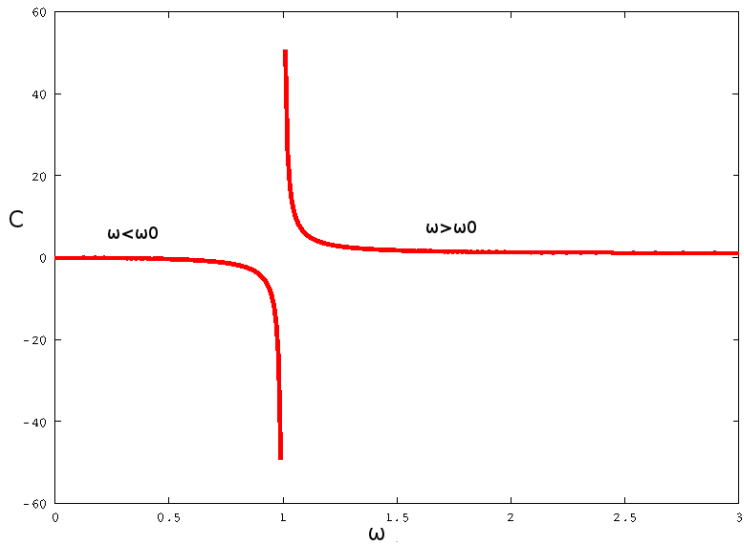}
\caption{(Color online) When the energy scale approches $\omega_{0}$ the a resonance condition appears to blowing up the divergences.}
\label{cf}
\end{figure}

Defining an interband energy scale for the difference of the Fermi points
\begin{eqnarray}
\omega_{0}=2v_{F}\Delta k_{F,R},
\end{eqnarray}
we can rewrite the cutoff function $C$, eq. (\ref{fuc}), independent of cutoff in the energy scale variable, as
\begin{eqnarray}
C(\omega, \omega_{0})&=& \frac{\omega^{2}}{\omega^{2} -\omega_{0}^{2}}.
\end{eqnarray} 
As such, this term drives to a resonance condition that blows up the divergences when energy scale variable 
approaches $\omega_{0}$. This behaviour is ilustrated in the figure \ref{cf}.
\begin{figure}[!htb]
\centering
\includegraphics[scale=0.6]{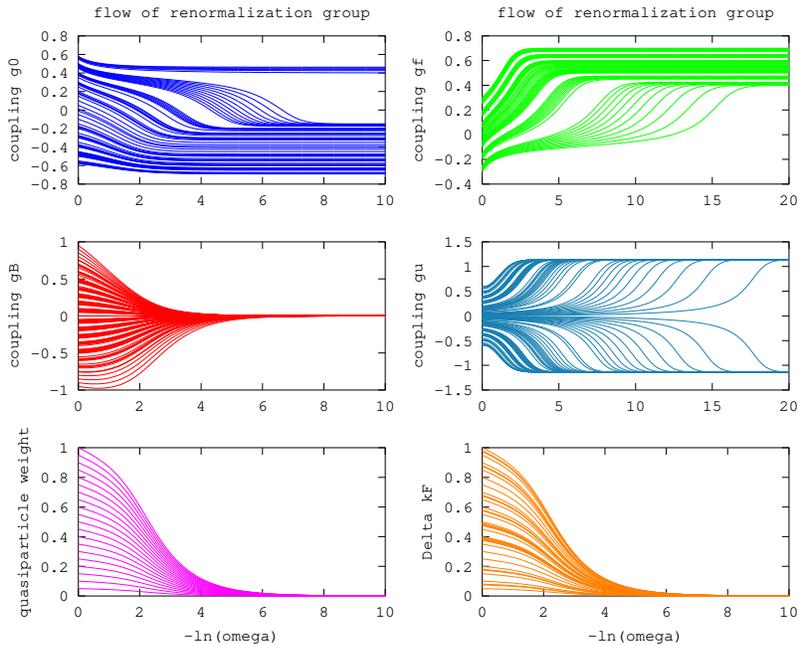}
\caption{(Color online) RG flows in the regime drived far from resonance, $\omega^{2}>>\omega_{0}^{2}$.}
\label{conf2neg45}
\end{figure}
\section{Confinement}

If the differences of Fermi points are driven in such a way to never be surpassed by the energy scale, in the 
effective relation $\omega^{2}>> \omega_{0}^{2}$, the cutoff contribution will be takes 
an effective form $C=1$, and the RG flow will be simplified 
in the numerical result displayed in the figure \ref{conf2neg45}, with an effective confinement when the energy scale associated to the 
interband differences of Fermi points never surpass the energy scale in the flow. Away from this condition, the resonance condition 
drives the system to a behaviour where confinement is not expected. 

\section{Conclusion}

We have shown that the recent cutoff dependence in the RG flow equations in the TCCM obtained in\cite{eberth} is 
a consequence of the use of the $l$ variable, 
leading to fixed points that are cutoff dependent. The numerical fixed points in terms of the $l$ variable are not universal, since 
they depend on the cutoff choice, as we have showed for the instances of cutoff choices $10$, $10^{2}$ and $10^{3}$. 
We have shown that this cutoff dependence can be simply removed by using the energy scale 
variable $\omega$. In this new form, the RG flow equations assume a cutoff independent form 
and consequently carry the universality property of RG invariance under cutoff transformations, leading to 
an unique set of fixed points independent of the cutoff choice. 

We have considered the cutoff function in the independent of cutoff form described by the energy scale variable and showed that 
this term crosses a resonance scale $\omega_{0}$ that blow up the divergences when the energy scale is driven towards $\omega_{0}$. 
By considering this problem, far from the resonance $\omega^{2}>>\omega_{0}^{2}$, a confinement behaviour of the interband Fermi 
points can be achieved in the fixed points in the RG flow. Away from this regime, the resonance condition 
drives the system away from confinement behaviour and consequetly the TCCM cannot merge in a Luttinger liquid phase.

\section{Acknowledgements}

T.P. thanks FAPEMA (Brazil)-APCInter 044/2013.

\end{document}